\documentclass[a4paper,11pt]{article}
\usepackage{pos}
\pdfoutput=1
\usepackage{algorithm}
\usepackage{algpseudocode}

\title{
Initial tensor construction for the tensor renormalization group}
\author*[a]{Katsumasa Nakayama}
\author[b]{Manuel Schneider}

\affiliation[a]{RIKEN Center for Computational Science, Kobe, 650-0047, Japan}

\affiliation[b]{Institute of Physics, National Yang Ming Chiao Tung University,\\
	1001 University Road, Hsinchu 30010, Taiwan}

\emailAdd{katsumasa.nakayama@riken.jp}

\abstract{We propose a method to construct the initial tensor representation of partition functions and observables for the tensor renormalization group (TRG). The TRG is a numerical calculation technique that utilizes a tensor network representations of physical quantities to investigate physical properties without encountering the sign problem.
To apply the TRG, it is essential to construct a locally connected tensor network suitable for recursive coarse-graining. We present a systematic approach for translating a general tensor representation of the partition function to this form. Furthermore, we show the dependence of TRG algorithms on the choice of the initial tensor network representation {and propose an improvement of TRG algorithms in this respect}.}

\FullConference{%
  The 41st International Symposium on Lattice Field Theory, LATTICE2024\\
  28th July - 3rd August 2024\\
  University of Liverpool, United Kingdom
}

\usepackage[english]{babel}
\usepackage[utf8]{inputenc}
\usepackage[T1]{fontenc}
\usepackage[babel,final]{microtype}
\usepackage{hyperref}			% links in document
\usepackage{url}
\usepackage{hyphsubst}		% hyphenation
\usepackage{float}				% place objects at the right position

\usepackage{amsfonts}
\usepackage{amsmath}
\usepackage{MnSymbol}	% \hateq for ^ over =
\usepackage{slashed}	% slashed operators
\usepackage{braket}

\usepackage[numbers,sort&compress]{natbib}
\usepackage{cleveref}
\crefname{section}{sec.}{sections}
\Crefname{section}{Section}{Sections}
\crefname{appendix}{app.}{appendices}
\Crefname{appendix}{Appendix}{Appendices}
\crefname{figure}{fig.}{figs.}
\Crefname{figure}{Figure}{Figures}
\crefname{equation}{eq.}{eqs.}
\Crefname{equation}{Equation}{Equations}

\usepackage[width=.9\textwidth,font=small,hypcap=true]{caption}
\usepackage[hypcap=true]{subcaption}

\def\be#1{\begin{equation}#1\end{equation}} %{equation}

\begin{document}
\maketitle

\section{Introduction}
Tensor network representations are widely used to calculate physical quantities~\cite{Meurice:2020pxc,Banuls:2022vxp}. If observables are expressed in the form of a tensor network, they can be calculated with tensor renormalization group (TRG) methods~\cite{TRG} that contract all indices of the tensor network. Since these TRG methods do not sample from a distribution function like Monte Carlo methods, they can be applied to systems that would otherwise suffer from a sign problem. Examples are systems at finite densities or with a topological charge~\cite{finitemu_TRG,CP1}. The TRG has been successfully applied to quantum field theories with gauge fields~\cite{Z2gauge,SU2SU3,initialTensorTRG,twocolorQCD}. Therefore, TRG methods provide a promising toolbox for studying high-energy physics where Monte Carlo cannot be applied.

The prerequisite for applying TRG methods is that the partition function or other observables are written in the form of a locally connected tensor network. In this proceeding, we introduce a generic method for finding such a representation. We start from the partition function that is expressed by the product of Boltzmann factors with a summation over physical degrees of freedom. In the two-dimensional Ising model, for example, the spin degrees of freedom are summed and the partition function can be expressed as
\begin{align}
    Z
        =&
        \sum_{\sigma =\pm 1}
        \prod_{x,y = 1} ^N
        e^{\frac{\beta g}{2}\sigma_{x,y}(\sigma_{x+1,y} + \sigma_{x,y+1})} %\nonumber\\
        =
        \sum_{\sigma =\pm 1}
        \prod_{x,y = 1} ^N
        K^\mathrm{(Ising)} _{\sigma_{x,y}\sigma_{x+1,y}\sigma_{x,y+1}}
    .\label{eq:Ising}
\end{align}
Here, $\beta$ is inverse temperature, $g$ is the coupling constant, and $\sigma$ are the spin variables at the lattice points $\{x,y\}$. The partition function is expressed as a product of the Boltzmann factors $K^\mathrm{(Ising)} _{\sigma_{x,y}\sigma_{x+1,y}\sigma_{x,y+1}}$. This is used as a starting point to construct a locally connected tensor network representation.

Note that the Boltzmann factor representation is not suitable for the application of TRG methods since it is not a tensor network. Although $K^\mathrm{(Ising)}$ are tensors of rank three, their indices are not contracted pairwise. A tensor network representation implies that an index that is summed over appears exactly twice in the product. We can therefore represent it graphically as a line that connects two tensors, or it could connect a tensor with itself. In the previous partition function, each index is shared by three tensors: $\sigma_{x,y}$ is an index of the tensors $K^\mathrm{(Ising)} _{\sigma_{x,y}\sigma_{x+1,y}\sigma_{x,y+1}}$, $K^\mathrm{(Ising)} _{\sigma_{x-1,y}\sigma_{x,y}\sigma_{x-1,y+1}}$, and $K^\mathrm{(Ising)} _{\sigma_{x,y-1}\sigma_{x+1,y-1}\sigma_{x,y}}$ in the product.

Common methods for constructing a tensor network representation use expansions such as the Taylor expansion, character expansion, or expansion by orthogonal functions~\cite{exp_const}.
For the example of the Ising model, a tensor network representation derived from a Taylor expansion~\cite{Ising_const,hotrg} is
\be{
    K^{(\mathrm{exp})}_{l_{x,y},l_{x+1,y},m_{x,y},m_{x,y+1}}	=
    	\sum_{\alpha}
    	W_{\alpha,l_{x,y}}
    	W_{\alpha,l_{x+1,y}}
    	W_{\alpha,m_{x,y}}
    	W_{\alpha,m_{x,y+1}},
     \label{eq:2d_ising_Kexp}
}
where the matrix $W$ is defined as
\be{
	W
		=
		\begin{pmatrix}
    		 \sqrt{\mathrm{cosh}(\beta g/2)} & {\sqrt{\mathrm{sinh}(\beta g/2)}} \\
    		 {\sqrt{\mathrm{cosh}(\beta g/2)}} & {-\sqrt{\mathrm{sinh}(\beta g/2)}}
        \end{pmatrix}.
    \label{eq:costFunction}
}

These constructions are model-specific and make use of certain properties of the Boltzmann weights. For example, the property $\sigma^2 = 1$ is used for the Ising model to find a finite-size tensor $K^\mathrm{(exp)}$. Consequently, these previous approaches cannot be generalized to arbitrary physical models. 
In this proceeding, we present an initial tensor construction method for general Boltzmann factor representations, without any expansions or decompositions.
The applicability of the method to gauge theories was shown for the $\mathbb{Z}_2$ model~\cite{initialTensorTRG}, where the free energy and specific heat were calculated in good agreement with previous TRG calculations and Monte Carlo simulations~\cite{Z2gauge}.

\section{Initial tensor construction for the Ising model}
\begin{figure}[htb]
    \centering
    \includegraphics[width=0.6\columnwidth, angle=0]{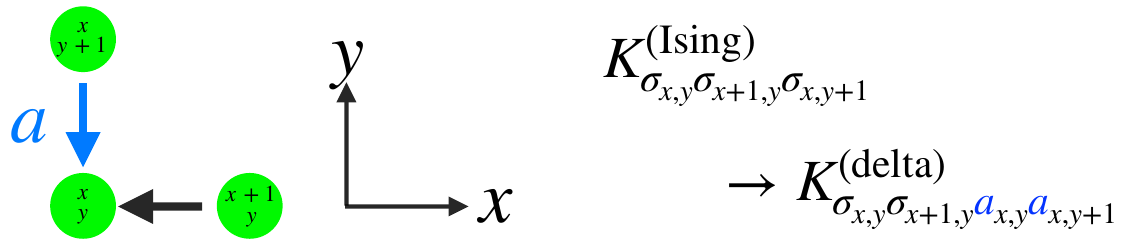}
    \caption{
        \label{fig:Ising}
        Schematic of the tensor network construction for the Ising model in the two dimensions. Green nodes represent the spin indices of a Boltzmann factor $K^\mathrm{(Ising)}$ in \cref{eq:Ising}. The blue vertical arrow represents the introduction of a new index $a_{x,y+1}$ as in \cref{eq:Ising_delta}, and a translation $a_{x,y+1} \rightarrow a_{x,y}$ as from \cref{eq:Ising_Z_delta} to \cref{eq:Ising_Z_shifted}. The black vertical arrow marks the nearest neighboring two unchanged spin variables in the resulting tensor $K^\mathrm{(delta)}$.
    }
\end{figure}

We explain our initial tensor construction for the Ising model with periodic boundary conditions in two-dimensional space-time in the following.
The starting point is the tensor representation of the partition function in terms of the Boltzmann factors $K^\mathrm{(Ising)} _{\sigma_{x,y}\sigma_{x+1,y}\sigma_{x,y+1}}$ from \cref{eq:Ising}.
We introduce a new index $a$ and use the identity
\be{
    K^\mathrm{(Ising)} _{\sigma_{x,y}\sigma_{x+1,y}\sigma_{x,y+1}}
        =
        \sum_{a = \pm 1}
        K^\mathrm{(Ising)} _{\sigma_{x,y}\sigma_{x+1,y}a_{x,y+1}}
        \delta_{\sigma_{x,y+1} a_{x,y+1}}.
        \label{eq:Ising_delta}
}
Note that this transformation does not require specific details of the Boltzmann factors $K$ or any numerical calculation.
The partition function becomes
\begin{align}
    Z
        =&
        \sum_{a,\sigma = \pm 1}
        \prod_{x,y = 1} ^N
        K^\mathrm{(Ising)} _{\sigma_{x,y}\sigma_{x+1,y}a_{x,y+1}} \delta_{\sigma_{x,y+1} a_{x,y+1}}
        \label{eq:Ising_Z_delta}\\
        =&
        \sum_{a,\sigma = \pm 1}
        \prod_{x,y = 1} ^N
        K^\mathrm{(Ising)} _{\sigma_{x,y}\sigma_{x+1,y}a_{x,y+1}} \delta_{\sigma_{x,y} a_{x,y}}
        . \label{eq:Ising_Z_shifted}
\end{align}
In the second line we shifted the indices of the delta functions by one site and made use of periodic boundary conditions.
With the definition of new tensors
\begin{equation}
    K^\mathrm{(delta)} _{\sigma_{x,y}\sigma_{x+1,y}a_{x,y}a_{x,y+1}}\equiv K^\mathrm{(Ising)} _{\sigma_{x,y}\sigma_{x+1,y}a_{x,y+1}} \times \delta_{\sigma_{x,y} a_{x,y}}
    \,\label{eq:Ising_initialTensor}
\end{equation}
the partition function can be written as
\be{
    Z
        =
        \sum_{a,\sigma = \pm 1}
        \prod_{x,y = 1} ^N
        K^\mathrm{(delta)} _{\sigma_{x,y}\sigma_{x+1,y}a_{x,y}a_{x,y+1}}.
}
This is a locally connected tensor network, since each index appears on exactly two nearest neighboring tensors $K^\mathrm{(delta)}$.
It is therefore a suitable starting point for the TRG method. \Cref{fig:Ising} shows a schematic picture of the index shift using a delta matrix for the Ising model.

\section{Initial tensor construction for general models}
For general models, the same index will not only be shared by three consecutive Boltzmann factors as in the two-dimensional Ising model. The initial tensor construction method can, however, be generalized.
The previous construction consisted of two steps:
\begin{enumerate}
    \item Insert a delta function, which creates a new index
    \item {Shift the delta function to a different site and combine it with the tensor at that site to create a new tensor that defines the partition function.}
\end{enumerate}
Although the partition function remains unchanged by these steps, the new tensor no longer depends on the index that was separated by the delta function.
The dependence was replaced by creating new variables. However, these new indices each appear on two neighboring tensors only, as required for a tensor contraction.

We formulate the concrete steps to construct a locally connected tensor network from any given Boltzmann weight as a pseudo-algorithm:
\begin{algorithm}
	\caption{Create a locally connected tensor network}
	\label{alg:initialTN}
	\begin{algorithmic}[1]
		\Require{Boltzmann weights $K$}
		\Ensure{Locally connected tensors for TRG}
		\State{Connect all sites that $K$ depends on by arrows; these should form a tree with arrows pointing from the roots to the origin}
        \label{alg:initialTN_SteinerTree}
		\While{arrows left in tree}
			\State{Pick a root with an arrow pointing from $\hat{r} - \hat{\mu}$ to  $\hat{r}$ in $\hat{\mu}$-direction}
            \label{alg:initialTN_pick}
            \State{Insert a delta function: $K_{\dots, \sigma_{\hat{r}-\hat{\mu}}} = \sum_{a_{\hat{v}}}  K_{\dots, a_{\hat{v}}} \delta_{\sigma_{\hat{r}-\hat{\mu}},a_{\hat{v}}}$}
            \label{alg:initialTN_delta}
            \State{Shift the delta function to a neighboring tensor: $K'_{\dots,\sigma_{\hat{r}} , a_{\hat{v}}, a_{\hat{v}+\hat{\mu}}} \equiv K_{\dots, a_{\hat{v}}} \delta_{\sigma_{\hat{r}}, a_{\hat{v}+\hat{\mu}}}$}
            \label{alg:initialTN_shift}
            \State{Replace $K$ by the new $K'$ and remove the arrow from the tree}
            \label{alg:initialTN_replace}
		\EndWhile
	\end{algorithmic}
\end{algorithm}

The newly introduced indices $a_{\hat{v}}$ have a site-index $\hat{v}$, which should be chosen such that the new tensor $K'^{(\hat{x})}_{\dots, a_{\hat{v}}, a_{\hat{v}+\hat{\mu}}}$ based at site $\hat{x}$ depends only on $a_{\hat{x}}$ and $a_{\hat{x} + |\hat{v}|}$.\footnote{In this convention, the final tensor will depend on indices at sites $(x,y)$ and $(x,y) + |\hat{\mu}|$ for all directions $\hat{\mu}$. Other conventions such as a dependence on $(x,y)$ and $(x,y) - |\hat{\mu}|$ would also be possible.} For two dimensions, this explicitly means:
\begin{itemize}
    \item[$\rightarrow$] Right arrow pointing from $(x'-1,y')$ to $(x',y')$ in direction $\hat{\mu} = \hat{x}$
    new index: $a_{\hat{v}} = a_{x,y}$\\
    decomposition: $K^{(x,y)}_{\dots, \sigma_{x'-1,y'}} = \sum_{a_{x,y}}  K_{\dots, a_{x,y}} \delta_{\sigma_{x'-1,y'}, a_{x,y}}$\\
    new tensor: $K'^{(x,y)}_{\dots, \sigma_{x',y'}, a_{x,y}, a_{x+1,y}} \equiv K^{(x,y)}_{\dots, a_{x,y}} \delta_{\sigma_{x',y'}, a_{x+1,y}}$
    \item[$\leftarrow$] Left arrow pointing from $(x'+1,y')$ to $(x',y')$ in direction $\hat{\mu} = -\hat{x}$;
    new index: $a_{\hat{v}} = a_{x+1,y}$\\
    decomposition: $K^{(x,y)}_{\dots, \sigma_{x'+1,y'}} = \sum_{a_{x+1,y}}  K_{\dots, a_{x+1,y}} \delta_{\sigma_{x'+1,y'}, a_{x+1,y}}$\\
    new tensor: $K'^{(x,y)}_{\dots, \sigma_{x',y'}, a_{x+1,y}, a_{x,y}} \equiv K^{(x,y)}_{\dots, a_{x+1,y}} \delta_{\sigma_{x',y'}, a_{x,y}}$
    \item[$\uparrow$] Up arrow pointing from $(x',y'-1)$ to $(x',y')$ in direction $\hat{\mu} = \hat{y}$;
    new index: $a_{\hat{v}} = a_{x,y}$\\
    new tensor: $K'^{(x,y)}_{\dots, \sigma_{x',y'}, a_{x,y}, a_{x,y+1}} \equiv K^{(x,y)}_{\dots, a_{x,y}} \delta_{\sigma_{x',y'}, a_{x,y+1}}$
    \item[$\downarrow$] Down arrow pointing from $(x',y'+1)$ to $(x',y')$ in direction $\hat{\mu} = -\hat{y}$;
    new index: $a_{\hat{v}} = a_{x,y+1}$\\
    new tensor: $K'^{(x,y)}_{\dots, \sigma_{x',y'}, a_{x,y+1}, a_{x,y}} \equiv K^{(x,y)}_{\dots, a_{x,y+1}} \delta_{\sigma_{x',y'}, a_{x,y}}$
\end{itemize}

We note that the steps in \cref{alg:initialTN_pick,alg:initialTN_delta,alg:initialTN_shift,alg:initialTN_replace} can be skipped for one final arrow pointing to the origin at $\hat{x} = (x,y)$. In diagrams, we draw the arrow pointing from $(x+1,y)$ to $(x,y)$ in black, indicating that we omit the transformation and are left with a dependence on the original indices $\sigma_{x,y}$ and $\sigma_{x+1,y}$.

In \cref{alg:initialTN_delta}, we label the new indices $a_{\hat{v}}$. In subsequent iterations, new names should be chosen, for example $a, b, c, \dots$.

\begin{figure}[htb]
    \centering
    \includegraphics[width=0.6\columnwidth, angle=0]{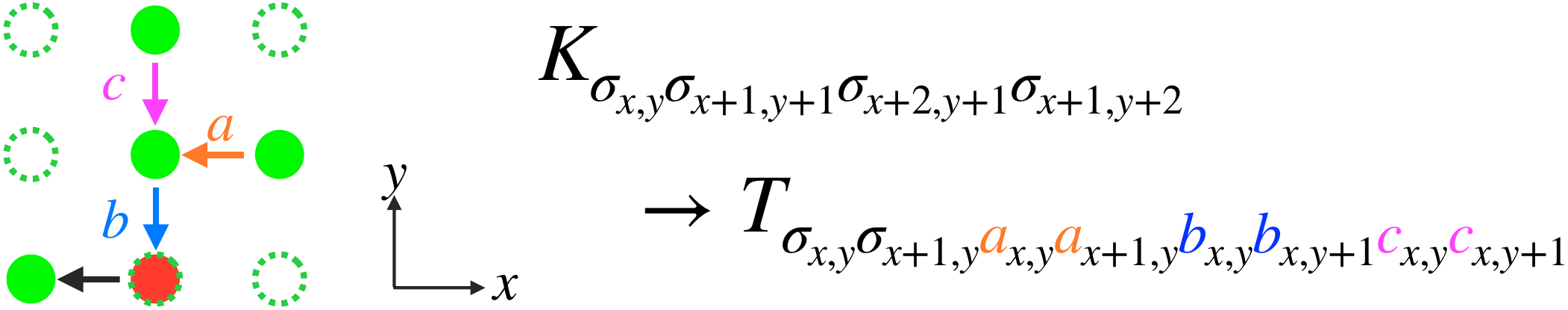}
    \caption{
        \label{fig:4-int}
        Schematic picture of the tensor network construction. The initial Boltzmann weight $K$ is transformed into tensors $T$, which are the building blocks of a locally connected tensor network for TRG coarse-graining. Filled green nodes represent the original degrees of freedom that $K$ depends on. Arrows are introduced to create a directed tree connecting all green nodes to the origin at $(x,y)$. Red nodes are additional points that need to be included in the tree. Each arrow introduces a new variable in $T$ according to \cref{alg:initialTN}. The final black arrow pointing to the origin leaves the spin variables unchanged.
    }
\end{figure}

To illustrate the procedure in more general cases, we discuss two additional examples. In the first case, we start from a Boltzmann factor representation which contains non-local interactions:
\be{
Z
=
\sum_{\sigma = \pm 1}
\prod_{x,y = 1} ^N
K_{\sigma_{x,y}\sigma_{x+1,y+1}\sigma_{x+2,y+1}\sigma_{x+1,y+2}}.
}
The tensor $K_{\sigma_{x,y}\sigma_{x+1,y+1}\sigma_{x+2,y+1}\sigma_{x+1,y+2}}$ has four different indices and can represent a variety of interaction terms~\cite{initialTensorTRG}. The diagram in \cref{fig:4-int} shows the directed tree that connects all sites involved in the interaction to the origin of the Boltzmann weight at $(x,y)$.

In order to construct the tensor network, we replace the arrow $\sigma_{x+2,y+1} \rightarrow \sigma_{x+1,y+1}$ and introduce the new index $a$. This is followed by a replacement of the arrow $\sigma_{x+1,y+1} \rightarrow \sigma_{x+1,y}$, which introduces the index $c$. In a last step, the arrow $\sigma_{x+1,y+2} \rightarrow \sigma_{x+1,y+1}$ creates index $b$.

The final tensor $T$, expressed in terms of the original Boltzmann weights $K$ becomes
\be{
T_{\sigma_{x,y}\sigma_{x+1,y}a_{x,y}a_{x+1,y}b_{x,y}b_{x,y+1}c_{x,y}c_{x,y+1}}
\equiv
K_{\sigma_{x,y}b_{x,y+1}a_{x+1,y}c_{x,y+1}}
\delta_{b_{x,y+1}a_{x,y}}
\delta_{b_{x,y+1}c_{x,y}}
\delta_{\sigma_{x+1,y}b_{x,y}}
.
}
Although the Boltzmann factor $K$ does not have an index $\sigma_{x+1,y}$, we have to add the point $(x+1,y)$ in the sequential index shifts to connect the other points to the origin. This is indicated by the red node in \cref{fig:4-int} which is part of the tree structure.

\begin{figure}[htb]
    \centering
    \includegraphics[width=0.6\columnwidth, angle=0]{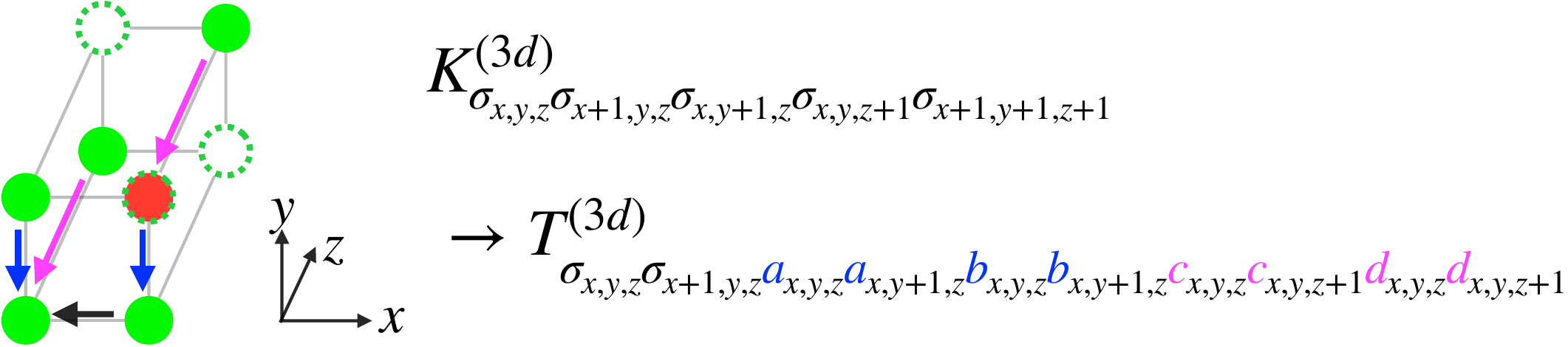}
    \caption{
        \label{fig:3d_steiner}
        Schematic picture of the tensor network construction in three dimensions. 
    }
\end{figure}

As a second example, we consider a three-dimensional system as depicted in \cref{fig:3d_steiner} with a Boltzmann factor
\be{
    Z^\mathrm{(3d)}
    =
    \sum_{\sigma = \pm 1}
    \prod_{x,y,z = 1} ^N
    K^\mathrm{(3d)} _{
        \sigma_{x,y,z}
        \sigma_{x+1,y,z}
        \sigma_{x,y+1,z}
        \sigma_{x,y,z+1}
        \sigma_{x+1,y+1,z+1}
    }.
}
The structure can incorporate nearest-neighbor interactions as well as an interaction on the diagonal of a cubic unit cell. The index shifts are indicated in \cref{fig:3d_steiner}. The final tensor $T$ has a larger size of $d^{10}$ if $d$ is the dimension of an initial degree of freedom $\sigma$. The resource scaling is discussed in more detail in~\cite{initialTensorTRG}.

The initial tensor construction does not assume any specific properties of the Boltzmann weights. Because of this, \cref{alg:initialTN} can be used in very general cases to construct the locally connected tensor network representation of a partition function.

\section{The Steiner tree problem}
For the initial tensor network construction according to \cref{alg:initialTN}, we have to connect all the sites on which the Boltzmann weight depends by a tree. This is the initial \cref{alg:initialTN_SteinerTree} of the algorithm and is depicted in \cref{fig:Ising,fig:4-int,fig:3d_steiner}. Every arrow in the tree increases the bond dimension of the final tensor. Therefore, the optimal choice minimizes the number of arrows in the tree. This is known as the rectilinear Steiner tree problem, which is a generalized minimum spanning tree problem~\cite{SpanningTrees}.
{The latter is the problem of finding line segments of minimum length which connect every given point.
The Steiner tree problem extends this by allowing one to add new points which also have to be connected by line segments. For square and cubic lattices we have the additional constraints that nodes must be on lattice points, and lines are only allowed between nearest neighbors.}

\begin{figure}[htb]
    \centering
    \includegraphics[width=0.6\columnwidth, angle=0]{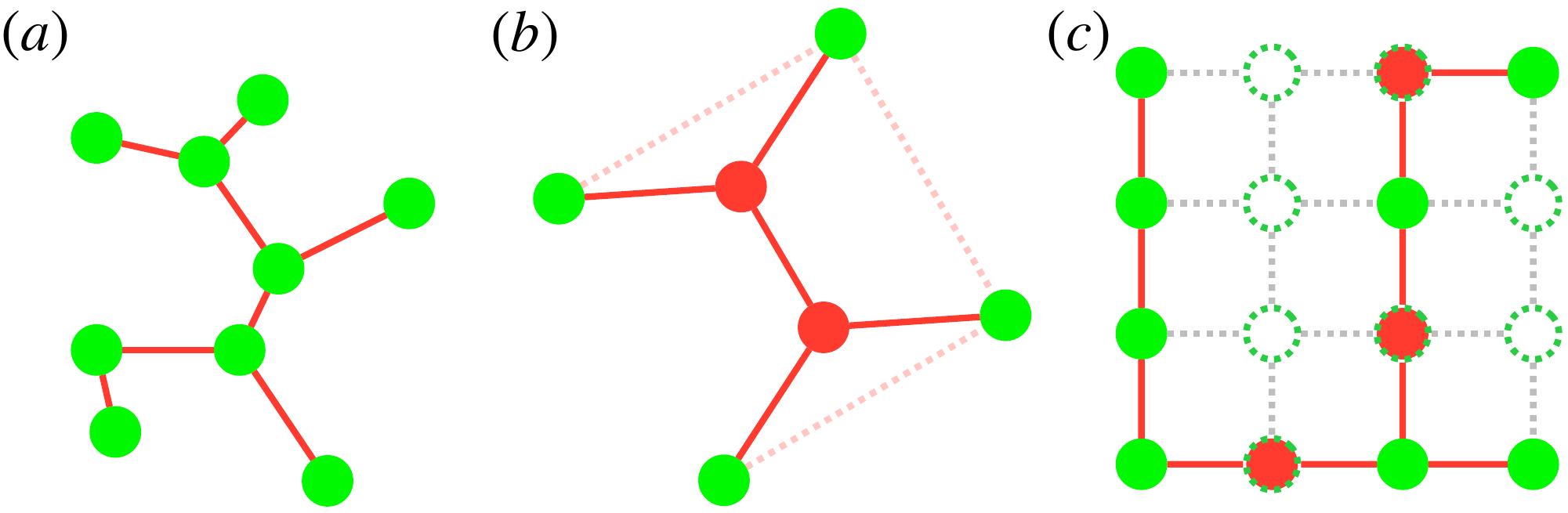}
    \caption{
        \label{fig:Steiner}
        Examples of (a) the minimum spanning tree problem, (b) the Steiner tree problem, and (c) the rectilinear Steiner tree problem. Green dots are the original nodes, and red dots are Steiner points that were added. {In (b), the dotted lines are the solution of the minimum spanning tree problem, which has a longer total line length than the Steiner tree solution.} 
        In (c), nodes and line segments are only allowed on a rectangular lattice of equal spacing (dotted nodes and lines).
    }
\end{figure}

\Cref{fig:Steiner} shows the minimum spanning tree and Steiner tree problems.
The Steiner tree problem is NP-hard, and the rectilinear Steiner tree problem is NP-complete.
Thus, the problem becomes very challenging if a large number of non-local interactions in the Boltzmann weights is given. 
However, relevant physical systems only include a small number of sites on which the Boltzmann weight depends. In these cases, the rectilinear Steiner problem can still be solved exactly by hand.

\section{Initial tensor dependence of tensor renormalization methods}
The partition function can be represented in many ways by different initial tensors. In this section, we use the example of the two-dimensional Ising model, where the tensors obtained from our method in \cref{eq:Ising_initialTensor} differ from the initial tensor in \cref{eq:2d_ising_Kexp}. We discuss how the accuracy of TRG algorithms with truncations depend on the form of the initial tensors. We suggest a way to remove this dependence by improvements using the boundary TRG method~\cite{boundarytrg,ATRG}.

\begin{figure}[h]
    \centering
    \includegraphics[width=0.5\columnwidth, angle=0]{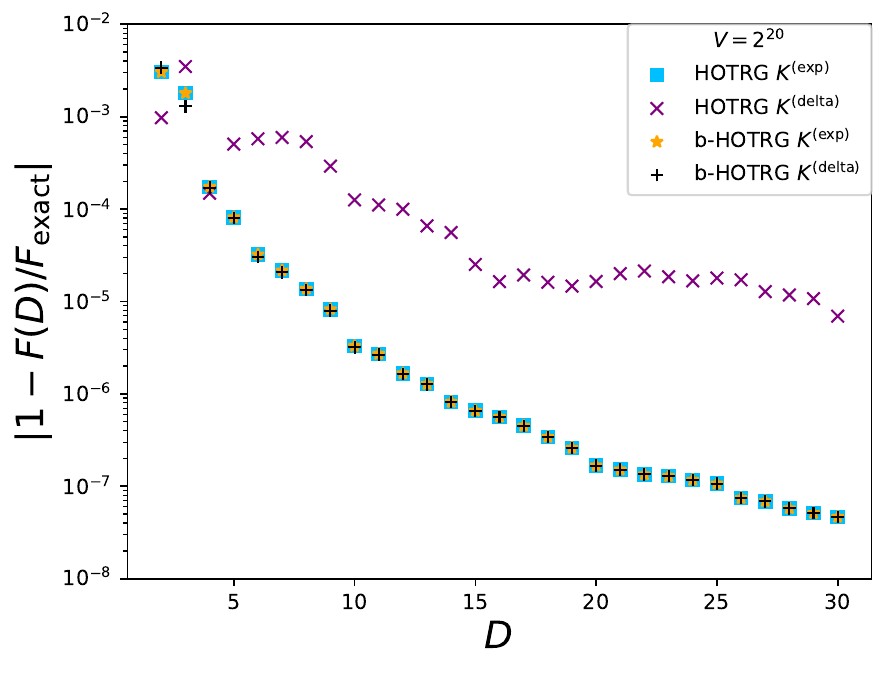}
    \caption{
        \label{fig:hotrg}
        Error of the free energy of the Ising model at the critical temperature from HOTRG and boundary-HOTRG. Different initial tensors are used to represent the partition function, see \cref{eq:2d_ising_Kexp,eq:Ising_initialTensor}.
    }
\end{figure}

We calculate the free energy
$
F
=
-
\frac{1}{\beta V}
\mathrm{ln}Z
$
for a volume $V=2^{20}$ at the critical temperature $\beta_c = \mathrm{ln}(1 + \sqrt{2})$ and compare to the exact solution~\cite{IsingExact}. We use the higher-order TRG (HOTRG) method~\cite{hotrg}. \Cref{fig:hotrg} shows the initial tensor dependence. The Taylor expansion method with initial tensors $K^\mathrm{(exp)}$ leads to better precision compared to our method with $K^\mathrm{(delta)}$.

\begin{figure}[htb]
    \centering
    \includegraphics[width=0.7\columnwidth, angle=0]{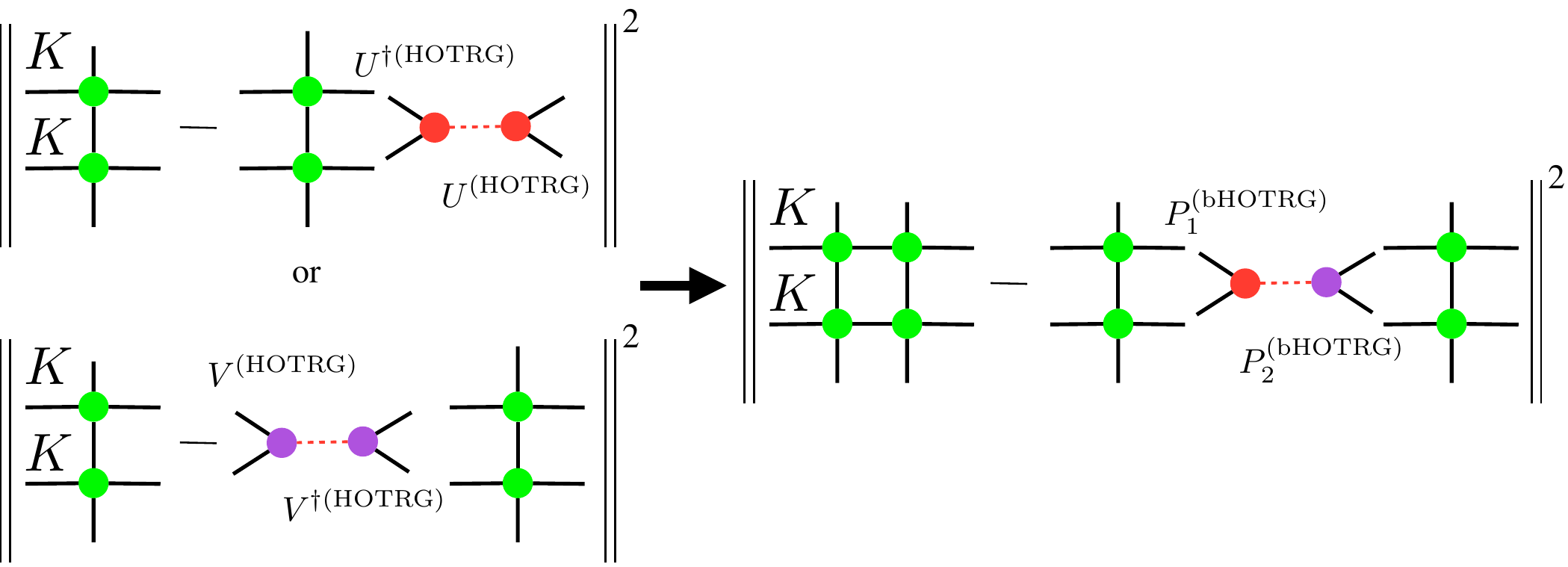}
    \caption{
        \label{fig:hotrg_to_bhotrg}
        The minimized cost function in the HOTRG (left) and boundary HOTRG (right).
    }
\end{figure}

In order to remove the dependence on the initial tensors, we apply the ideas of the boundary TRG method~\cite{boundarytrg} to HOTRG. All TRG methods introduce a truncation, which minimizes a cost function. The cost functions of HOTRG and boundary HOTRG are shown in \cref{fig:hotrg_to_bhotrg}. While HOTRG uses only the left or right isometries of a truncated singular value decomposition (SVD), the boundary HOTRG method includes both of them in so called squeezers $P$ which are used for the truncation~\cite{ATRG}. This also corresponds to a larger cell that is taken into account in the cost function.

Specifically, the isometry $U^\mathrm{(HOTRG)}$ is obtained by a truncated SVD of $KK$ with singular values $(\lambda^{(U)})$. It connects to the right indices of the product $KK$ and can be used to truncate the indices and create coarse-grained tensors. Similarly, the isometry $V^\mathrm{(HOTRG)}$ truncates the left indices, with corresponding singular values $(\lambda^{(V)})$. HOTRG uses either $U$ or $V$ for the coarse-graining and thus introduces an asymmetry in the truncation step. This is the reason why HOTRG depends strongly on the symmetry properties of the initial tensors~\cite{initialTensorTRG}. In the boundary HOTRG method, we instead take into account both $U^\mathrm{(HOTRG)}$ and $V^\mathrm{(HOTRG)}$. For this, we consider the additional SVD $\lambda^{(U)}U^\mathrm{(HOTRG)}V^\mathrm{(HOTRG)}\lambda^{(V)} = U\Lambda V$.
The squeezers $P_1 ^\mathrm{(bHOTRG)}$ and $P_2 ^\mathrm{(bHOTRG)}$ that are used in the truncation are then defined as
$P_1 ^\mathrm{(bHOTRG)}\equiv V^\mathrm{(HOTRG)}\lambda^{(V)} V^\dagger / \sqrt{\Lambda}$,
$P_2 ^\mathrm{(bHOTRG)}\equiv  (1/ \sqrt{\Lambda}) U^\dagger \lambda^{(U)} U^\mathrm{(HOTRG)}.
$
Further details can be found in~\cite{initialTensorTRG}.

When the boundary HOTRG method is applied, the initial tensors $K^\mathrm{(delta)}$ and $K^\mathrm{(exp)}$ lead to the same accuracy of the free energy, as shown in \cref{fig:hotrg}. The boundary HOTRG method can reproduce the previous result of HOTRG with $K^\mathrm{(exp)}$. Our initial tensor construction thus leads to similarly good results as previous, problem-specific methods. In addition, the boundary TRG improvement is not only applicable to HOTRG but can be used in any TRG method~\cite{initialTensorTRG,ATRG,TriadTRG,RHOTRG}. Replacing isometries by squeezers removes the dependence on the form of the initial tensors.

\section{Conclusions}
We have shown a general method to find the tensor network representation of a partition function. These initial tensors can be used in tensor renormalization group algorithms. Even though we assumed translational-invariant systems, the method is applicable to inhomogeneous systems. It can, thus, be used not only for the partition function but for general observables, which can include impurity tensors.

The initial tensor construction starts from the Boltzmann weights and replaces original indices by newly introduced ones, which are shifted in the direction of a given path. This path is given as a solution of the Steiner tree problem. Although the latter problem is difficult to solve in general, we can typically find the solution easily for physically relevant systems with a limited number of non-local interactions.
The solution of the Steiner tree problem directly relates to the size of the initial tensors, which is important for the accuracy of the coarse-graining. Thus, we can estimate the applicability and numerical demands of TRG methods in advance.

We further discussed the dependence of TRG methods on the form of the initial tensors. Algorithms like HOTRG lead to inaccurate results for non-symmetric initial tensors. However, this dependence can be removed by introducing squeezers in the coarse-graining step, similarly to the boundary TRG method. In this way, common TRG methods can be improved, and the initial tensor construction presented in this work leads to accuracies similar to those of previous constructions.

We assumed periodic boundary conditions in order to move the delta functions from one tensor to its nearest neighbor. The method can, however, also be used for other boundary conditions. In this case, the boundary tensors differ from the tensors in the bulk.

\acknowledgments
This work is based on the publication~\cite{initialTensorTRG} and was supported by JSPS KAKENHI Grant No. 24K17059, and Taiwanese NSTC Grant No. 113-2119-M-007-013.

\bibliography{bibliography}

\end{document}